\documentclass[aps,superscriptaddress,twocolumn,a4paper,showpacs]{revtex4}
\usepackage{graphicx}
\usepackage{amsmath}
\usepackage{amssymb}
\usepackage{enumerate}
\usepackage{subfigure}
\usepackage{tabularx}
\usepackage[colorlinks=true, pdfstartview=FitV, linkcolor=blue, citecolor=red, urlcolor=black, breaklinks=true]{hyperref}
\usepackage[all]{xy}
\usepackage{amssymb}
\usepackage{color}
\newcommand{\be}{\begin{equation}}
\newcommand{\ee}{\end{equation}}
\newcommand{\ben}{\begin{eqnarray}}
\newcommand{\een}{\end{eqnarray}}
\newcommand{\bes}{\begin{subequations}}
\newcommand{\ees}{\end{subequations}}
\newcommand{\wt}{\widetilde}

\newcommand{\bb}{\bibitem}

\newcommand{\LL}{{\cal L}}

\newcommand{\vphi}{{\varphi}}
\newcommand{\veps}{{\varepsilon}}

\begin{document}
%%%%%%%%%%%%%%%%%%%%%%%%%%%%%%%%%%%%%%
\title{Twinlike Models for Kinks, Vortices and Monopoles}
\author{D. Bazeia}\email{bazeia@fisica.ufpb.br}\affiliation{Departamento de F\'\i sica, Universidade Federal da Para\'\i ba, 58051-970 Jo\~ao Pessoa, PB, Brazil}
\author{M.A. Marques}\email{mam.matheus@gmail.com}\affiliation{Departamento de F\'\i sica, Universidade Federal da Para\'\i ba, 58051-970 Jo\~ao Pessoa, PB, Brazil}
\author{R. Menezes}\email{rmenezes@dce.ufpb.br}
\affiliation{Departamento de Ci\^encias Exatas, Universidade Federal
da Para\'{\i}ba, 58297-000 Rio Tinto, PB, Brazil}
\affiliation{Departamento de F\'\i sica, Universidade Federal de Campina Grande, 58109-970, Campina Grande, PB, Brazil}\pacs{}
\date{\today}
\begin{abstract}
This work deals with twinlike models that support topological structures such as kinks, vortices and monopoles. We investigate the equations of motion and develop the first order framework to show how to build distinct models with the same solution and energy density, as required to make them twinlike models. We also investigate how the stability under small fluctuations behaves and introduce the conditions to get the same stability on general grounds. In particular, we study models that support kinks, vortices and monopoles in one, two, and three spatial dimensions, respectively.
\end{abstract}
\maketitle

%%%%%%%%%%%%%%%%%%%%%%%
\section{Introduction}
In high energy physics, topological defects are static solutions of the equations of motion that appear in a diversity of contexts \cite{sut,vilenkin}. The simplest topological defects are kinks \cite{vachaspati}, which appear in the presence of scalar fields in $(1,1)$ spacetime dimensions. They may also arise when a complex scalar field is minimally coupled with an Abelian gauge field in $(1,2)$ spacetime dimensions, and here they are called vortices \cite{no}. Furthermore, one can study a theory in $(1,3)$ spacetime dimensions with scalar fields and non-Abelian gauge fields under the $SU(2)$ symmetry to find monopoles \cite{thooft,polyakov}.

The above topological structures appear in models that engender standard kinematics, but one can also deal with generalized models. An interesting example of this appeared in Ref.~\cite{kinf} in the context of Cosmology. Later, other generalized models were used to study the cosmic coincidence problem \cite{cosm1,cosm2}. These non-canonical models present different features as, for instance, the capacity of driving inflation without the presence of potential terms. 

Along the same lines, other possibilities appeared in Refs.~\cite{babichev1,babichev2}, in which defect structures were studied in models with generalized kinematics. Over the last years, similar generalizations have been studied in several papers \cite{kd1,kd2,kd3,kd4,kd5,kd6,kd7,kd8}, with distinct motivation.

As a particularly interesting issue, in Ref.~\cite{trodden} it was found that generalized models may support the same topological configurations as their standard counterparts. Since these models present the same solution, with the same energy density, they were called twinlike models. However, their linear stability spectra are not equal in general, except in some special cases discussed before in Refs.~\cite{twin2,les1,les2}. Since the work \cite{trodden}, twinlike models for kinks have been studied by several authors  \cite{twin2,les1,les2,twin1,twin3,twin4,twin5,twin6}. The cases of twinlike vortices and monopoles are more complicated, and only very recently an investigation on twinlike vortices has been carried out in Ref.~\cite{twinvortex}.

Generalized models of the form that we consider in this work have been studied in many other contexts, in particular as tachyon dynamics and condensation \cite{S,S0,S1,T1,T2,D,T3}, as tachyonic dark energy in cosmology \cite{pad,baz,berto}, as cosmological constraints on tachyon dark energy models \cite{P}, as non-canonical fields in braneworld scenarios with an extra dimension of infinite extent \cite{db1,db2}, as the Eddington-Inspired Born-Infeld braneworld scenario \cite{C}, in Born-Infeld/gravity correspondence \cite{G} and in Born-Infeld gravity \cite{big1,big2,big3,big4,RN}. They are also of current interest to study holographic entanglement entropy in Born-Infeld eletrodynamics \cite{HE}, and in other scenarios.

The purpose of this work is to study twinlike topological defects of the three distinct types mentioned above. We follow the natural order and start the investigation in the next Sec.~\ref{skinks}, where one reviews the basic properties of twinlike kinks, and add new results on their stability. In Sec.~\ref{svortices} we study twinlike vortices, and there one shows explicitly the conditions to make the generalized and canonical model become twins. Also, we investigate the stability and find conditions to have vortices with the same stability of the canonical case, up to a given order in the fluctuations. In Sec.~\ref{smonopoles} we investigate the harder case of monopoles, exhibiting the general conditions for the existence of twinlike models having the same stability spectra. Finally, we present our comments and conclusions in Sec.~\ref{conclusions}.

%%%%%%%%%%%%%%%%%%%%%%
\section{Kinks}\label{skinks}

The case of kinks is the simplest one, and we take the generalized action for a single real scalar field with the metric obeying $\eta_{\mu\nu} = \text{diag}(+,-)$. It has the form
\be\label{gactionkink}
S=-\int d^2x\,V(\phi)G(Z),
\ee
with $V(\phi)$ being the potential and $G(Z)$ describing an arbitrary function of $Z$, which is defined as 
\be\label{Xkink}
Z = -\frac{X}{V(\phi)} \quad \text{and} \quad X=\frac12 \partial_\mu\phi\partial^\mu\phi.
\ee
The standard case is obtained for $G(Z) = 1+Z$, which gives the action
\be\label{sactionkink}
S_s=\int d^2x\,\left(X-V(\phi)\right).
\ee One can vary the action \eqref{gactionkink} with respect to the scalar field to get the equation of motion
\be\label{eomkink}
\partial_\mu(G_Z\partial^\mu \phi) = (ZG_Z-G)V_\phi,
\ee
with $G_Z=dG/dZ$, $V_\phi=dV/d\phi$, and so on. To search for kinklike solutions, we consider static configurations, that is, we take $\phi=\phi(x)$. In this case, from Eq.~\eqref{Xkink} one gets $X=-{\phi^\prime}^2/2$. The equation of motion \eqref{eomkink} assumes the form
\be\label{eomk}
\left( G_Z \phi^\prime\right)^\prime = (G-ZG_Z)V_\phi.
\ee
Here one takes the boundary conditions $\phi(\pm\infty)=v_\pm$, where $v_\pm$ are minima of the potential, which is supposed to describe spontaneous symmetry breaking. 

Invariance under spacetime translations $x_\mu\to x_\mu^\prime$, with $x^\prime_\mu=x_\mu+a_\mu$ leads to the energy-momentum tensor $T^{\mu\nu}$ with components $T^{00}$ as the energy density and $T^{11}$ as the stress component.  The energy density is given by
\be\label{rhokink}
\rho = V(\phi) G(Z),
\ee
and, in the case of stressless solutions, the above Eq.~\eqref{eomk} can be integrated to give the first order equation
\be\label{fogkink}
{\phi^\prime}^2 = \frac{G}{G_Z} V(\phi).
\ee
By using the definition of $Z$ in Eq.~\eqref{Xkink}, we get
\be\label{gzeqkink}
G-2ZG_Z=0.
\ee 
We emphasize that the above Eq.~\eqref{gzeqkink} is an algebraic equation for $Z$. As the action \eqref{gactionkink} shows, one has to make $G(Z)$ explicit to define the model; thus, one uses $G$ and $G_Z$ in \eqref{gzeqkink} to solve it for $Z$. Since we are interested in real configurations, the solution of \eqref{gzeqkink} has to be a real constant, which we call $Z=c^2$, for convenience. This, combined with Eqs.~\eqref{Xkink}, leads to the first order equation
\be\label{fokink}
\frac12 {\phi^\prime}^2 = c^2 V(\phi).
\ee
Then, one can show that the solution is given by $\phi(x) = \phi_s(\tilde{x})$, where $\tilde{x}=c x$ and $\phi_s(x)$ is the solution for the standard case, with $G=1+Z$. From Eq.~\eqref{rhokink}, we get the energy density
\be\label{edenstkink}
\rho(x) = \frac{G(c^2)}{2} \rho_s(\tilde{x}),
\ee
where $\rho_s(x) = 2V(\phi_s(x))$, and the energy
\be
E = G(c^2)\int_{-\infty}^\infty dx V(\phi(\tilde{x})) = \frac{G(c^2)}{2c}E_s,
\ee
where $E_s$ is the energy of the standard case. Thus, we see that the model presents the solution of the standard case if $c=1$, and the energy density of the standard case if $G(1)=2$. Also, we get $G_Z(1)=1$ from Eq.~\eqref{gzeqkink}. Then, the generalized models \eqref{gactionkink} and the standard model $G(Z)=1+Z$ as in Eq.~\eqref{sactionkink} are twins if one imposes the conditions
\be\label{tcondkink}
c=1,\quad G(1)=2, \quad\text{and}\quad G_Z(1)=1.
\ee

\subsection{Stability}

We now study the stability of the twinlike solutions by considering fluctuations of the field in the model. We then write $\phi(x,t) = \wt{\phi}(x) + \eta(x,t)$ in the action \eqref{gactionkink} up to the second order in $\eta(x,t)$. We are taking $\wt{\phi}(x)$ as the static topological solutions of \eqref{fokink}. This makes $X$ in Eq.~\eqref{Xkink} to assume the form $X=X_0 + X_1 + X_2$, where $X_i$ contains the dependence of $X$ in the $i$-th order of $\eta$, as given below:
\be\label{Xetakink}
X_0 =\frac12 \partial_\mu\wt{\phi}\partial^\mu\wt{\phi},\quad X_1 = \partial_\mu\wt{\phi}\partial^\mu\eta, \quad X_2 =\frac12 \partial_\mu\eta\partial^\mu\eta.
\ee
We only write the second order contribution of $\eta$ in the action \eqref{gactionkink} because the first order terms vanishes with the use of the equation of motion \eqref{eomkink}. It has the form $S^{(2)} = \int d^2x \LL^{(2)}$, where
\be\label{l2kink}
\begin{split}
\LL^{(2)}=&\;G_Z X_2 -\frac{G_{ZZ}}{2V}(X_1)^2 -\frac{V_\phi}{V} ZG_{ZZ} \eta X_1 \\
 &- \left(V_{\phi\phi}(G-ZG_Z) + Z^2G_{ZZ}\frac{V_\phi^2}{V} \right)\frac{\eta^2}{2}.
\end{split}
\ee
In the standard case, $G(Z)=1+Z$, we get
\be\label{sl2kink}
\LL_s^{(2)}= X_2 -V_{\phi\phi}\frac{\eta^2}{2}.
\ee
As one can see, the conditions \eqref{tcondkink} in Eq.~\eqref{l2kink} lead to
\be
\LL^{(2)}=X_2- V_{\phi\phi}\frac{\eta^2}{2} -\frac{G_{ZZ}}{2V}\left(X_1 +{V_\phi}\eta\right)^2.
\ee
Thus, the above expression is not equal to Eq.~\eqref{sl2kink} because the $G_{ZZ}$ terms do not vanish. However, if we want the contribution $S^{(2)}$, which drives the stability of the system, to be equal to the one of the standard case up to second order in $\eta$, we have to impose an additional condition to $G(Z)$, given by $G_{ZZ}(1)=0$.

If one consider contributions of $\eta$ up to order ${N}$, the action \eqref{gactionkink} can be written as
\be
S = \sum_{p=2}^{N} S^{(p)},
\ee
where $S^{(p)}$ contains the $p$-th order contributions of $\eta$ in the action. However, for the conditions $c=1$, $G(1)=2$ and $G_Z(1)=1$, these contributions are not equal to the one for the standard case. It is possible to show that $S^{(p)}$ becomes equal to the one of the standard case if all the derivatives from the second order up to the $p$-th order of $G(Z)$ at $Z=1$ are null, ie., for
\be\label{testcondkink}
\left.\frac{d^2 G}{dZ^2}\right|_{Z=1} = \left.\frac{d^3 G}{dZ^3}\right|_{Z=1} = \ldots = \left.\frac{d^p G}{dZ^p}\right|_{Z=1} = 0.
\ee
This peculiar characteristic appears only for solutions whose stress component of the energy-momentum tensor is null, since this is the only case in which Eq.~\eqref{gzeqkink} holds. Note that the order of the expansion, ${N}$, informs how indistinguishable are the stability spectra of the two models, because the conditions \eqref{tcondkink} and \eqref{testcondkink} give the same solutions, energy densities and stability up to order ${N}$. If ${N}$ increases, the models tend to present the same physical features, at least in the case concerning fluctuations around the static solution. Evidently, the maximum value of $N$ which is allowed depends on the specific forms of $V(\phi)$ and $G(Z)$ that are used to define the generalized model.

We illustrate the general results with the standard model with $G(Z)=1+Z$, and the generalized model with $G=1+Z+(1-Z)^k$, for $k=2,3,4...$\,. For this generalized model one also gets from \eqref{gzeqkink} that $Z=1$ and $G(1)=2$ and $G_Z(1)=1$. Thus, the above generalized model is twin of the standard model. Also, the generalized model has the very same stability behavior, until order $N$ for $N=k-1$ and $k=3,4,...$\,.

%%%%%%%%%%%%%%%%%%%%%%%%%%%%%%%%
\section{Vortices}\label{svortices}

Let us now consider the action for a gauge field and a complex scalar field in the $(1,2)$ flat spacetime, with metric $\eta_{\mu\nu}={\rm diag}(+,-,-)$. In this case, it is possible to extend the previous formalism and propose the action
\be\label{gactionvortex}
S=\int d^3x\,\Big(X-V(|\vphi|)G(Z)\Big),
\ee
with $V$ being the potential and
\be\label{XYvortex}
Z=-\frac{Y}{V}, \quad X= \overline{D_\mu \vphi}D^\mu \vphi, \quad \text{and}\quad Y=-\frac14 F_{\mu\nu}F^{\mu\nu},
\ee
where $D_\mu = \partial_\mu +ieA_\mu$, $F_{\mu\nu}=\partial_\mu A_\nu - \partial_\nu A_\mu$ and the overline stands for the complex conjugation. The standard case is obtained for $G(Z)=1+Z$, with action
\be\label{sactionvortex}
S_s=\int d^3x\,\Big(X+Y-V(|\vphi|)\Big).
\ee
The variation of the action \eqref{gactionvortex} with respect to the fields $\overline\vphi$ and $A_\mu$ gives the equations of motion
\bes\label{geomvortex}
\ben
 D_\mu D^\mu\vphi&=& \frac{\vphi}{2|\vphi|}(ZG_Z-G)V_{|\vphi|}, \\ \label{meqsvortex}
 \partial_\mu \left(G_Z F^{\mu\nu} \right) &=& J^\nu ,
\een
\ees
where the current is $J_\mu = ie(\bar{\vphi}D_\mu \vphi-\vphi\overline{D_\mu\vphi})$. In the case of static solutions, we can consider the gauge $A_0=0$ since the temporal component of Eq.~\eqref{meqsvortex}, which is Gauss' law for our model, is compatible with this condition. This makes the vortex electrically neutral since  $(J^0=0)$ and so it carries on electrical charge. We define $B = -F^{12}$, which makes the functions $X$ and $Y$ in Eq.~\eqref{XYvortex} become
\be\label{XYstaticvortex}
X = -\overline{D_i \vphi}D_i \vphi \quad \text{and} \quad Y= -\frac12 B^2.
\ee
The energy density is given by
\be\label{grhovortex}
\rho = -X+VG(Z).
\ee
We take the usual ansatz for vortices
\bes\label{ansatzvortex}
\ben
\vphi(r,\theta)&=&g(r)e^{i n\theta},\\
A_i&=&-\epsilon_{ij} \frac{x^j}{er^2}[a(r)-n],
\een
\ees
where $r$ and $\theta$ are polar coordinates and $n$ the vortex winding number.  The functions $g(r)$ and $a(r)$ must obey the boundary conditions
\bes\label{bcondvortex}
\begin{align}
g(0)&=0, & a(0)&=n, \\ 
\lim_{r\to\infty}{g(r)}&=v, & \lim_{r\to\infty}{a(r)}&=0.
\end{align}
\ees
Here, $v$ is the parameter associated to the symmetry breaking. With this ansatz, the functions $X$ and $Y$ in Eq.~\eqref{XYstaticvortex} becomes
\be\label{XYansatzvortex}
X=-\left({g^\prime}^2+\frac{a^2g^2}{r^2}\right) \quad \text{and} \quad Y=-\frac{{a^\prime}^2}{2e^2r^2},
\ee
where the prime denotes derivative with respect to $r$. Furthermore, the magnetic field assumes the form $B=-F^{12}=-a^\prime/(er)$ and the magnetic flux $\Phi=2\pi\int_0^\infty r dr B(r)$ is quantized:
\be\label{mflux}
\Phi=\frac{2\pi n}{e}.
\ee
The equations of motion \eqref{geomvortex} with the ansatz \eqref{ansatzvortex} become
\bes\label{secansatzvortex}
\begin{align}
\frac{1}{r} \left(rg^\prime\right)^\prime -\frac{a^2g}{r^2} + \frac12 (ZG_Z-G)V_{|\vphi|} &= 0, \\
r\left(G_Z\frac{a^\prime}{r} \right)^\prime - 2e^2ag^2 &= 0.
\end{align}
\ees
In the standard case, $G=1+Z$, the energy density from Eq.~\eqref{grhovortex} is
\be\label{rhosv}
\begin{split}
\rho &= -X-Y+V(|\vphi|) \\
     &={g^\prime}^2 + \frac{a^2g^2}{r^2} + \frac{{a^\prime}^2}{2e^2r^2} + V(g).
\end{split}
\ee
If the potential is 
\be\label{vsvortex}
V_s(|\vphi|)=e^2(v^2-|\vphi|^2)^2/2,
\ee
the BPS formalism \cite{ps,bogo} allows showing that the first order equations 
\be\label{fostandardvortex}
{g_s^\prime} =  \frac{a_sg_s}{r} \quad\text{and}\quad \frac{{a_s^\prime}}{er} =-\sqrt{2V_s(g_s)},
\ee
lead to the equations of motion \eqref{secansatzvortex}. Their analitical solutions are unkown. We can put these first order equations in the energy density \eqref{rhosv} to get
\be
\rho_s=2{g_s^\prime}^2+\frac{{a_s^\prime}^2}{e^2r^2}.
\ee
One can integrate the above expression in the space and define the gradient and magnetic energy respectively as
\be\label{energyportions}
E_{sg}=4\pi\int_0^\infty r dr {g_s^\prime}^2 \quad\text{and}\quad E_{sm} = \frac{\pi}{e^2}\int_0^\infty dr  \frac{{a_s^\prime}^2}{r},
\ee
to get the total energy as $E=E_{sg}+2E_{sm}$. The energy is quantized and it is given by $E=2\pi |n| v^2$.

For a general $G(Z)$, we can consider the first-order equations 
\bes\label{fovortex}
\begin{align}
{g^\prime} &=  \frac{ag}{r}, \\ \label{foavortex}
\frac{{a^\prime}}{er} &= - \sqrt{\frac{G V_s(g)}{G_Z}}.
\end{align}
\ees
To use these equations in the context of the action \eqref{gactionvortex}, we combine them with the expression for $Y$ in Eq.~\eqref{XYansatzvortex} and use the definition of $Z$ in Eq.~\eqref{XYvortex} to get the equation
\be\label{gzeqvortex}
G-2ZG_Z=0.
\ee
The issue here is similar to the one discussed for Eq.~\eqref{gzeqkink} in the previous section. This is an algebraic equation that must be solved for $Z$, and the solution is $Z=constant$, which we write as $Z=c^4$, for convenience. We can use this in Eqs.~\eqref{fovortex} to get
\bes\label{fovortexc}
\begin{align}
{g^\prime} &=  \frac{ag}{r}, \\ \label{foavortexc}
\frac{{a^\prime}}{er} &= - c^2\sqrt{2 V_s(g)}.
\end{align}
\ees
One can show that the solutions of Eqs.~\eqref{fovortexc} are $a(r)=a_s(\tilde{r})$ and $g(r)=g_s(\tilde{r})$, where $a_s$ and $g_s$ are the solutions of Eqs.~\eqref{fostandardvortex} for the standard case and $\tilde{r}=c\,r$. We now check if Eqs.~\eqref{fovortexc} are compatible with the equations of motion. After substituting \eqref{fovortexc} into \eqref{secansatzvortex}, we get that
\be\label{constvortex}
c^2G_Z(c^4) = 1.
\ee
The above algebraic equation constrains the value of the constant $c$. Finally, Eqs.~\eqref{fovortex} with Eq.~\eqref{XYansatzvortex} allow us to write the energy density \eqref{grhovortex} in this case as
\be\label{edenstvortex}
\begin{split}
\rho &=2{g^\prime}^2+G_Z\frac{{a^\prime}^2}{e^2r^2} \\
     &=2c^2\left(\frac{d g_s(\tilde{r})}{d\tilde{r}}\right)^2 + \frac{G_Z c^4}{e^2\tilde{r}^2}\left(\frac{d a_s(\tilde{r})}{d\tilde{r}}\right)^2.
\end{split}
\ee
We integrate the above expression to get the energy $E = E_{sg} + 2E_{sm}G_Zc^2$, where $E_{sg}$ and
$E_{sm}$ are the gradient and magnetic energies as in Eq.~\eqref{energyportions} for the standard model. Therefore, we see that $c=1$ gives exactly the solutions of the standard case and $G_Z(1)=1$ gives exactly the standard energy density. Also, from Eq.~\eqref{gzeqvortex} we get $G(1)=2$. These conditions are all compatible with the constraint \eqref{constvortex}. Thus, \textit{in summa}, if the conditions
\be\label{tcondvortex}
c=1, \quad G(1)=2, \quad\text{and}\quad G_Z(1)=1,
\ee
are satisfied, the class of models \eqref{gactionvortex} and the standard model $G(Z)=1+Z$ from Eq.~\eqref{sactionvortex} are twins. Although twinlike vortices were presented before in Ref.~\cite{twin6}, the above conditions are originally presented in the current work.

\subsection{Stability}

We now study the stability of the twinlike solutions by considering fluctuations of the scalar and gauge fields in our model. We then write $\vphi(\vec{r},t) =\wt{\vphi}(\vec{r}) + \eta(\vec{r},t)$ and $A_\mu(\vec{r},t) = \wt{A}_{\mu}(\vec{r}) + \xi_\mu(\vec{r},t)$ in the action \eqref{gactionvortex} up to the second order in $\eta(\vec{r},t)$ and $\xi_\mu(\vec{r},t)$. The tilde stands for the static solutions of Eq.~\eqref{fovortex}. We only write the second order contribution in the action because the first order terms vanishes with the use of the equations of motion \eqref{geomvortex}. It has the form $S^{(2)} = \int d^3x \LL^{(2)}$, where
\be\label{l2vortex}
\begin{split}
\LL^{(2)} =&\; X_2 +G_Z Y_2 - \frac{G_{ZZ}}{2V} (Y_1)^2 -\frac{V_{|\vphi|}}{|\wt{\vphi}|V} ZG_{ZZ}Y_1\Re(\overline{\wt{\vphi}}\,\eta)  \\
 &+\frac18\left[(G-ZG_Z)\left(\frac{V_{|\vphi|}}{|\wt{\vphi}|} - V_{|\vphi||\vphi|}  \right) - ZG_{ZZ} \frac{V_{|\vphi|}}{V}  \right] \\
 &\times\left[|\eta|^2+\frac{2}{|\wt{\vphi}|^2}\Re(\overline{\wt{\vphi}}^2\eta^2) \right].
\end{split}
\ee
Here, $\Re(z)$ represents the real part of $z$. The functions $X_i$ and $Y_i$ stand for the $i$-th order contributions of the fluctuations of the functions $X$ and $Y$ present in Eq.~\eqref{XYvortex}, whose expression may be written as $X=\sum_{i=1}^4 X_i$ and $Y=\sum_{i=1}^2 Y_i$. Since we are considering contributions up to second order in the fluctuations, we can neglect $X_3$ and $X_4$ in the expansion. The relevant terms are
\bes
\begin{align}
X_0 &= \overline{\wt{D_\mu \vphi}}\wt{D^\mu \vphi}, \\
X_1 &= 2\Re\left(\overline{\wt{D_\mu \vphi}}\left(\partial^\mu\eta + ie(\wt{A}^\mu\eta+\xi^\mu\wt{\vphi})\right)\right), \\
X_2 &= \left|\partial_\mu\eta + ie(\wt{A}_\mu\eta+\xi_\mu\wt{\vphi})\right|^2+ 2e\Re\left(i\overline{\wt{D_\mu \vphi}}\xi^\mu\eta\right), \\
Y_0 &= -\frac14 \wt{F}_{\mu\nu}\wt{F}^{\mu\nu}, \\
Y_1 &= -\frac12 \wt{F}_{\mu\nu}(\partial^\mu\xi^\nu-\partial^\nu\xi^\mu), \\
Y_2 &= -\frac14 (\partial_\mu\xi_\nu-\partial_\nu\xi_\mu)(\partial^\mu\xi^\nu-\partial^\nu\xi^\mu).
\end{align}
\ees
For the standard case, $G(Z)=1+Z$, we get from Eq.~\eqref{l2vortex} that
\be\label{sl2vortex}
\begin{split}
\LL_s^{(2)} =&\; X_2\! +\! Y_2 +\! \frac18\!\left(\frac{V_{|\vphi|}}{|\vphi|} - V_{|\vphi||\vphi|}\!\right)\!\left(|\eta|^2\!+\!\frac{2}{|\wt{\vphi}|^2}\Re(\overline{\wt{\vphi}}^2\eta^2) \right).
\end{split}
\ee
By using the conditions \eqref{tcondvortex} in Eq.~\eqref{l2vortex} we get
\be\label{l2vortex1}
\begin{split}
\LL^{(2)} =&\; X_2\! +\!Y_2 + \!\frac18\!\left(\frac{V_{|\vphi|}}{|\vphi|} - V_{|\vphi||\vphi|} \! \right)\!\left(|\eta|^2\!+\!\frac{2}{|\wt{\vphi}|^2}\Re(\overline{\wt{\vphi}}^2\eta^2)\!\right) \\
&- \frac{G_{ZZ}}{2V}\left[ (Y_1)^2 +\frac{2V_{|\vphi|}}{|\wt{\vphi}|} Y_1\Re(\overline{\wt{\vphi}}\,\eta)+\right.  \\
 &\left.+\frac18  {V_{|\vphi|}} \left(|\eta|^2+\frac{2}{|\wt{\vphi}|^2}\Re(\overline{\wt{\vphi}}^2\eta^2) \right)\right].
\end{split}
\ee
Thus, the above expression it is not equal to Eq.~\eqref{sl2vortex} because the $G_{ZZ}$ terms do not vanish. This is similar to what happened for kinks in the previous section. If we want to have the same stability of the standard case up to second order in $\eta$ and $\xi_\mu$, we have to impose the aditional condition $G_{ZZ}=0$.

If one consider contributions of $\eta$ and $\xi_\mu$ up to order ${N}$, one has to take $X_3$ and $X_4$ in account and the action \eqref{gactionvortex} can be written as
\be
S = \sum_{p=2}^{{N}} S^{(p)},
\ee
where $S^{(p)}$ contains the $p$-th order contributions of $\eta$ and $\xi_\mu$ in the action. However, for the conditions \eqref{tcondvortex}, these contributions are not equal to the one for the standard case. As was done in the previous section, it is possible to show that $S^{(p)}$ becomes equal to the one of the standard case if all the derivatives from the second order up to the $p$-th order of $G(Z)$ at $Z=1$ are null, ie., for
\be
\left.\frac{d^2 G}{dZ^2}\right|_{Z=1} = \left.\frac{d^3 G}{dZ^3}\right|_{Z=1} = \ldots = \left.\frac{d^p G}{dZ^p}\right|_{Z=1} = 0.
\ee
As in previous section, we remark here that the stability spectra of the models tend to become indistinguishable as ${N}$ increases to larger and larger values.

%%%%%%%%%%%%%%%%%%%%%%%%%%%%%%%%%%%%%
\section{Monopoles}\label{smonopoles}
In this section, we start with the $SU(2)$ Yang-Mills-Higgs generalized action
\be\label{gactionmonopole}
S=\int d^4x\,\Big(Y G(Z) - V(|\phi|)\Big),
\ee
with $V$ being the potential and
\be\label{XYgeralmonopole}
Z=\frac{X}{Y},\! \quad X=-\frac12 D_\mu \phi^a D^\mu \phi^a,\! \quad \text{and}\! \quad Y=-\frac{1}{4}F^{a}_{\mu\nu}F^{a\mu\nu},
\ee
where $|\phi|=\sqrt{\phi^a\phi^a}$, $D_\mu \phi^a = \partial_\mu\phi^a + g\veps^{abc}A^b_\mu\phi^c$ and $F^a_{\mu\nu} = \partial_\mu A^a_\nu - \partial_\nu A^a_\mu + g\veps^{abc}A^b_\mu A^c_\nu$. Here, $g$ is the coupling constant, the indices $a,b,c=1,2,3$ stand for the $SU(2)$ symmetry of the fields and the greek letters $\mu,\nu=0,1,2,3$ represent the spacetime indices. For convenience, we use the metric tensor as $\eta_{\mu\nu} = \textrm{diag}(-,+,+,+)$, which is different from the one used in the previous sections for kinks and vortices.

We note that the action \eqref{gactionmonopole} is now different, with the definitions \eqref{XYgeralmonopole} do not involving the potential $V$. The issue here is that in the case of monopoles, the standard model develops interesting topological structures in the absence of potential \cite{ps,bogo}, so we had to work hard to build the above action, controlled by Eqs.~\eqref{gactionmonopole} and \eqref{XYgeralmonopole}.

It is straightforward to show that $G(Z)=1+Z$ gives the standard case with the action
\be\label{sactionmonopole}
S=\int d^4x\,\Big(X+Y - V(|\phi|)\Big).
\ee
The equations of motion associated to the action \eqref{gactionmonopole} are given by
\bes\label{eomgeral}
\begin{align}
D_\mu\left(G_Z D^\mu \phi^a\right) &= \frac{\phi^a}{|\phi|} V_{|\phi|}, \\
D_\mu\left((G-ZG_Z) F^{a\mu\nu}\right) &= G_Z g\veps^{abc}\phi^b D^\nu \phi^c.
\end{align}
\ees

We now work on static configurations with $A^a_0=0$, since we want to study magnetic monopoles. We define $B^a_i = \frac{1}{2}\veps_{ijk}F^{ajk}$, which makes the functions $X$ and $Y$ in Eq.~\eqref{XYgeralmonopole} become
\be\label{XYstatic}
X = -\frac12 D_i\phi^a D_i\phi^a \quad \text{and} \quad Y= -\frac12 B^a_i B^a_i.
\ee
The energy density for the model under the above conditions is given by
\be\label{rhomonopole}
\rho = V-YG(Z).
\ee

We now take the usual ansatz for monopoles
\be\label{hedgehog}
\phi^a = \frac{x_a}{r} H(r) \quad \text{and} \quad A_i^a = \veps_{abi}\frac{x_b}{gr^2}(K(r)-1),
\ee
with the boundary conditions
\bes\label{bcondmonopole}
\begin{align}
H(0)&=0, & K(0)&=1, \\
\lim_{r\to\infty}{H(r)} &= \pm v, & \lim_{r\to\infty}{K(r)} &= 0.
\end{align}
\ees
The parameter $v$ is involved in the symmetry breaking which we suppose is present in the potential $V$. One can show that the equations of motion \eqref{eomgeral} assume the form
\bes\label{eomansatzmonopole}
\begin{align}
\left(r^2 G_Z H^\prime\right)^\prime &= 2G_Z H K^2 + r^2 V_{|\phi|},   \\
\begin{split}
r^2\left((G-ZG_Z) K^\prime\right)^\prime &= K\left(G_Z g^2r^2H^2 + \right. \\
                                         & \hspace{4mm} \left. +(G-ZG_Z)(K^2-1)\right) ,
\end{split}
\end{align}
\ees
and the functions $X$ and $Y$ in Eqs.~\eqref{XYstatic} become
\bes\label{XYmonopoleansatz}
\begin{align}
X &=-\frac12 \left({H^\prime}^2 + \frac{2H^2K^2}{r^2} \right),\\
Y &= -\frac12 \left(\frac{2{K^\prime}^2}{g^2r^2}+\frac{(1-K^2)^2}{g^2r^4} \right).
\end{align}
\ees
In the standard case, $G=1+Z$, it is known from the BPS formalism \cite{ps,bogo} that for $V(|\phi|)=0$, the equations of motion \eqref{eomansatzmonopole} can be integrated to give
\be\label{firstorderstandardmonopole}
gr^2H_s^\prime = \pm (1-K_s^2) \quad\text{and}\quad K_s^\prime = \mp gH_sK_s,
\ee
whose solutions are known as
\bes\label{solstandardmonopole}
\begin{align}
H_s(r) &=\pm\left(v\coth(vgr)-\frac{1}{gr}\right), \\
K_s(r) &=v gr\,\text{csch}(v gr).
\end{align}
\ees
In this case, the energy density \eqref{rhomonopole} is given by
\be\label{rhostandardmonopole}
\begin{split}
\rho_s &= X+Y\\
       &= \frac{g^2v^4(1+2\cosh^2(v g r))}{\sinh^4(v g r)}  - \frac{4 gv^3\cosh(v g r)}{r\sinh^3(v g r)} + \frac{1}{g^2 r^4}.
\end{split}
\ee
The energy is given by $E_s=4\pi v/g$.

One follows this route and sets $V=0$ for a general $G(Z)$. We find that the equations of motion \eqref{eomansatzmonopole} can be integrated to give the first order equations
\bes\label{firstordermonopole}
\begin{align}
\sqrt{G_Z}gr^2H^\prime &= \pm \sqrt{G-ZG_Z}(1-K^2), \\
\sqrt{G-ZG_Z}K^\prime &= \mp \sqrt{G_Z}gHK,
\end{align}
\ees
for constant $Z$, without loss of generality. In fact, we use the above equations and Eqs.~\eqref{XYmonopoleansatz} to find the algebraic equation
\be\label{gzeqmonopole}
G-2ZG_Z=0,
\ee
which has to be solved for $Z$, as it also appeared before in the case of kinks and vortices. We set $Z=c^2$, where $c$ is a real parameter which is squared for convenience. This can be used into Eqs.~\eqref{firstordermonopole}, which assume the form
\be\label{fomonopole}
gr^2H^\prime = \pm c(1-K^2) \quad\text{and}\quad cK^\prime = \mp gHK.
\ee
Thus, the solutions are $K(r)=K_s(r)$ and $H(r)=c H_s(r)$, where $K_s(r)$ and $H_s(r)$ are given by Eqs.~\eqref{solstandardmonopole}. From Eq.~\eqref{rhomonopole}, one can show that the energy density is given by
\be\label{edensmonopole}
\rho = \frac{G\left(c^2\right)}{2}\rho_s,
\ee
where $\rho_s$ is given as in Eq.~\eqref{rhostandardmonopole}. Thus, the energy is related to the one of the standard case, in the form $E=E_sG(c^2)/2$. Therefore, Eqs.~\eqref{fomonopole} gives the standard solutions \eqref{solstandardmonopole} if $c=1$ and the energy density \eqref{edensmonopole} reproduces Eq.~\eqref{rhostandardmonopole} if $G(1)=2$. Thus, if the conditions
\be\label{tcondmonopole}
c=1, \quad G(1)=2, \quad\text{and}\quad G_Z(1)=1,
\ee
are satisfied, the class of models \eqref{gactionmonopole} and the standard model $G(Z)=1+Z$ are twinlike models.

\subsection{Stability}

We now study the stability of the twinlike solutions by considering fluctiations of the scalar and gauge fields in our model. We then write $\phi^a(\vec{r},t) = \wt{\phi}^a(\vec{r}) + \eta^a(\vec{r},t)$ and $A^a_\mu(\vec{r},t) = \wt{A}^a_{\mu}(\vec{r}) + \xi^a_\mu(\vec{r},t)$ in the action \eqref{gactionmonopole} up to the second order in $\eta(\vec{r},t)$ and $\xi_\mu(\vec{r})$. The tilde stands for the solutions of Eqs.~\eqref{eomansatzmonopole}. We only write the second order contribution of the fluctuations in the action because the first order terms vanishes with the use of the equations of motion \eqref{eomgeral}. It has the form $S^{(2)} = \int d^4 x \LL^{(2)}$, where
\be\label{l2monopole}
\begin{split}
\LL^{(2)} =&\; G_Z X_2 +(G-ZG_Z) Y_2 + \frac{G_{ZZ}}{2Y} (X_1)^2  \\
 &- \frac{Z^2G_{ZZ}}{2Y}(Y_1)^2-\frac{ZG_{ZZ}}{Y}X_1 Y_1 \\
 & +\frac12\left(\frac{V_{|\phi|}}{|\wt{\phi}|} - V_{|\phi||\phi|}\right)\frac{\wt{\phi}^a\wt{\phi}^b}{|\wt{\phi}|^2}\eta^a\eta^b.
\end{split}
\ee
The functions $X_i$ and $Y_i$ represent the $i$-th order contributions of the fluctuations of the functions $X$ and $Y$ present in Eq.~\eqref{XYgeralmonopole}, whose expression can be written as $X=\sum_{i=1}^4 X_i$ and $Y=\sum_{i=1}^4 Y_i$. Since we are considering contributions up to second order in the fluctuations, we can neglect $X_3$, $X_4$, $Y_3$ and $Y_4$ in the respective expansions. The relevant terms are
\bes
\begin{align}
X_0 &= -\frac12 \wt{D_\mu\phi^a} \wt{D^\mu \phi^a}, \\
X_1 &= -\wt{D^\mu \phi^a}\left(\partial_\mu\eta^a+g\veps^{abc}(\wt{A}^b_\mu\eta^c+\xi^b_\mu\wt{\phi}^c)\right), \\
\begin{split}
X_2 &= -\frac12\left|\partial_\mu\eta^a+g\veps^{abc}(\wt{A}^b_\mu\eta^c+\xi^b_\mu\wt{\phi}^c)\right|^2 \\
&\hspace{4mm}-\!g\veps^{abc}\wt{D^\mu \phi^a}\xi_\mu^b\eta^c,
\end{split}
\\
Y_0 &= -\frac14 \wt{F}^a_{\mu\nu}\wt{F}^{a\mu\nu}, \\
Y_1 &= -\frac12 \wt{F}^{a\mu\nu}\left(\partial_\mu\xi^a_\nu-\partial_\nu\xi^a_\mu+g\veps^{abc}(\wt{A}^b_\mu\xi^c_\nu + \xi^b_\mu\wt{A}^c_\nu)\right), \\
\begin{split}
Y_2 &= -\frac14 \left|\partial_\mu\xi^a_\nu-\partial_\nu\xi^a_\mu+g\veps^{abc}(\wt{A}^b_\mu\xi^c_\nu + \xi^b_\mu\wt{A}^c_\nu)\right|^2\\
&\hspace{4mm}-\!\frac12 g\veps^{abc}\wt{F}^a_{\mu\nu}\xi^{b\mu}\xi^{c\nu}.
\end{split}
\end{align}
\ees
One can see that the potential $V(|\phi|)$ is present in Eq.~\eqref{l2monopole}. However, to get twin models we set $V(|\phi|)=0$. In this case, Eq.~\eqref{l2monopole} simplifies to
\be\label{l2v0monopole}
\begin{split}
\LL^{(2)} =&\; G_Z X_2 +(G-ZG_Z) Y_2 + \frac{G_{ZZ}}{2Y} (X_1)^2  \\
 &- \frac{Z^2G_{ZZ}}{2Y}(Y_1)^2-\frac{ZG_{ZZ}}{Y}X_1 Y_1.
\end{split}
\ee
For the standard case, $G=1+Z$, we have
\be\label{l2smonopole}
\LL_s^{(2)} =X_2 +Y_2.
\ee
By using the conditions \eqref{tcondmonopole} in Eq.~\eqref{l2v0monopole} we get
\be
\LL^{(2)} = X_2 + Y_2 + \frac{G_{ZZ}}{2Y}\left(X_1+Y_1\right)^2,
\ee
which is not equal to Eq.~\eqref{l2smonopole}, as in the cases of kinks and vortices studied in the previous sections. The result is similar: if we want to have the same stability of the standard case up to second order in $\eta^a$ and $\xi^a_\mu$, we have to impose the additional condition $G_{ZZ}=0$.

If one consider contributions of $\eta^a$ and $\xi^a_\mu$ up to order ${N}$, one has to take $X_3$, $X_4$, $Y_3$ and $Y_4$ into account and the action \eqref{gactionmonopole} can be written as
\be
S = \sum_{p=2}^{{N}} S^{(p)},
\ee
where $S^{(p)}$ stand for the $p$-th order contributions of $\eta^a$ and $\xi^a_\mu$ in the action. However, for the conditions \eqref{tcondmonopole}, these contributions are not equal to the one for the standard case. As was done in the previous sections, it is possible to show that $S^{(p)}$ becomes equal to the one of the standard case if all the derivatives from the second order up to the $p$-th order of $G(Z)$ at $Z=1$ are null, ie., for 
\be
\left.\frac{d^2 G}{dZ^2}\right|_{Z=1} = \left.\frac{d^3 G}{dZ^3}\right|_{Z=1} = \ldots = \left.\frac{d^p G}{dZ^p}\right|_{Z=1} = 0.
\ee
Here, as done in the previous sections, the stability spectra of the models tend to become indistinguishable as ${N}$ gets bigger and bigger.

%%%%%%%%%%%%%%%%%%%%%%%%%%%%%%%%%%%%%%%%%
\section{Comments and conclusions}\label{conclusions}

In this work we have studied twinlike models, i.e., models described by distinct Lagrange densities, but having the same solutions and energy density. We investigated the cases of kinks, vortices and monopoles in two, three, and four spacetime dimensions, respectively. They arise in generalized models that obey similar conditions, although $Z$ represents different quantities in each one of the three distinct cases. Another distinction appears in the energy density \eqref{edenstvortex} for vortices, which assumes a dependence on the function $G(Z)$ that is different from the two other cases of kinks \eqref{edenstkink} and monopoles \eqref{bcondmonopole}. 

Other interesting results concern stability of kinks, vortices and monopoles. We have investigated how the stability behaves in each case and have found conditions to get models having the stability spectrum as indistinguishable as we want. 

Furthermore, it is worth commenting that, although we have investigated twinlike models when one considers the standard model and another one, generalized, it is possible to extend this formalism to find a generalized model which is twin to another generalized model. To illustrate this one considers the following action for vortices
\be
S=\int d^3x\Big(X-V(|\vphi|)G(Z)\Big),
\ee
where
\be 
Z=-\frac{P(|\vphi|)F_{\mu\nu}F^{\mu\nu}}{4V(|\vphi|)},
\ee
with $P(|\vphi|)$ being an arbitrary dimensionless function and $X=|D_\mu\vphi|^2$. In this case, we inform that it is possible to find conditions that make the above generalized model twin to another generalized model, in the form
\be
S = \int d^3x\Big(X-\frac{P(|\vphi|)}{4}F_{\mu\nu}F^{\mu\nu} - V(|\vphi|)\Big).
\ee
This line of investigation is part of another work, with focus on generalized twinlike models, with a generalized model being twin to another generalized model. 

As we have shown explicitly in this work, there are several examples of kinks, vortices and monopoles, which appear as solutions of standard and generalized theories that engender the same physical features, having the same stability spectra up to a given order ${N}$.\\

%%%%%%%%%%%%%%%%%%%%%%%%%%%%%%%%%%%%%%%%%%%%%%%%%%%
\acknowledgements{The authors acknowledge the Brazilian agency CNPq for partial financial support. DB thanks support from the contracts CNPq:455931/2014-3 and CNPq:306614/2014-6, MAM thanks support from the contract CNPq:140735/2015-1, and RM thanks support from the contracts CNPq:455619/2014-0 and CNPq:306826/2015-1.}

%%%%%%%%%%%%%%%%%%%%%%%%%%%%%%%%%%%%%%%%%%%%%%%%%%%

%%%%%%%%%%%%%%%%%%%%%%%%%%%%%%%%%%%%%%%%%%%%%%%%%%%
\end{document}